\documentclass[twocolumn]{aastex701}
\shorttitle{Shocks in small radio sources}
\shortauthors{Lopez \& Lacy}

\accepted{October 25, 2026}
\submitjournal{ApJL}

\begin{document}

\title{Evidence for strong shocks in the ISM of small radio sources}

\author[orcid=0009-0009-0998-5049,sname=Lopez]{Valeria Lopez}
\affiliation{National Radio Astronomy Observatory, 520 Edgemont Road, Charlottesville VA 22903}
\affiliation{Center for Astrophysics $|$ Harvard \& Smithsonian, 60 Garden Street, Cambridge, MA 02138, USA}
\email[show]{valelop800@gmail.com}

\author[orcid=0000-0002-3032-1783,gname=Mark, sname=Lacy]{Mark Lacy}
\affiliation{National Radio Astronomy Observatory, 520 Edgemont Road, Charlottesville VA 22903}
\email[show]{mlacy@nrao.edu}

\begin{abstract}
Radio jets deposit energy into the interstellar medium of their host galaxies, and while a jet is still confined within the host it is expected to drive shocks through the gas. Broad [O~{\sc iii}]~$\lambda$5007 profiles trace ionized gas that has been disturbed, the [O~{\sc i}]/H$\alpha$ versus [O~{\sc iii}]/H$\beta$ diagram helps distinguish gas photoionized by an AGN from gas excited by shocks, and the high ionization [Ne~{\sc v}]~$\lambda$3426 line traces the hard UV continuum of the accretion disk, which shocks only reproduce at very high velocities. We have selected two samples of radio-loud AGN from a cross-match of the VLA Sky Survey with DESI Data Release 1, at $0.15<z<0.5$ and with [O~{\sc i}]~$\lambda$6300 detected at $>5\sigma$: one with projected radio sizes of 10--15~kpc, whose jets are still inside the host, and one with sizes greater than 25~kpc, whose jets have broken out. We fit the [O~{\sc iii}] and [O~{\sc i}] profiles, measure W80, and classify each source on the BPT diagram. The confined sources have broader [O~{\sc iii}] lines, sit systematically closer to the LINER region, and are detected less often in [Ne~{\sc v}], and the [Ne~{\sc v}] difference is not explained by redshift or dust. The broad components are not systematically blueshifted, so the confined sources are distinguished by line width rather than by outflow velocity. We then discuss how these results indicate that the jets of the confined sources are driving strong shocks through the interstellar medium of their hosts.
\end{abstract}


\section{Introduction}

The radio jets launched by active galactic nuclei (AGN) are among the most energetic structures produced by galaxies. At the center of an active galaxy sits a supermassive black hole, and when that black hole is accreting, it can drive collimated outflows of plasma outward through the galaxy. These jets pass through the interstellar medium of the host, heating the gas, and, in the most powerful cases, removing it from the galaxy altogether. This link between galaxy evolution and AGN activity is called feedback. Without feedback, models of galaxy evolution cannot reproduce the mass distribution of galaxies that we see today
\citep{croton06,heckman2014}.

Ionized gas is one of the more convenient ways to watch feedback happen. The [O~{\sc iii}]~$\lambda$5007 line is bright in the narrow-line regions of AGN and has been used for a long time to trace gas that has been set into motion by the AGN \citep[e.g.][]{whittle1985}. Both the shape of this line and its flux carry information about the gas. A common measure of the shape is W80, which is the velocity range that holds the middle 80 percent of the flux in the line \citep[e.g.][]{liu2013}. Narrow values of W80 mean the gas dynamics reflect the virial motions within the gravitational potential well of the galaxy. Broad values, above roughly 800~km~s$^{-1}$, mean there is gas at velocities that rotation by itself cannot account for, which is what an outflow looks like in a spectrum. A broad profile by itself does not say what disturbed the gas, however. Radiation from the nucleus, winds, and shocks caused by radio jets can all move the ionized gas, and separating these possibilities requires comparing the line widths against some property of the AGN itself. In this work that property is the size of the radio source.

Simulations \citep[e.g.][]{mukherjee25} indicate that radio jets will have a strong effect on the ISM and observations have shown consistency with this. 
\citet{mullaney2013} show that radio emitting AGN with compact structures have systematically broader [O~{\sc iii}] emission lines than radio-quiet AGN in stacked spectra from the Sloan Digital Sky Survey (SDSS). \citet{molyneux2019} classified 2922 $z<0.2$ SDSS AGN as radio compact or extended and found that the compact sources have systematically broader [O~{\sc iii}] profiles.
Broad components with FWHM~$>1000$~km~s$^{-1}$, the signature of extreme outflows, were twice as common in the compact sources overall, and nearly four times as common in the moderate radio luminosity range where the AGN dominates the radio emission.
Their interpretation is that young or frustrated jets, still confined within the host, are responsible for the most disturbed gas. \citet{escott25} found that AGN detected in the LOFAR Two-metre Sky Survey were more likely to show outflows in their [O~{\sc iii}]~$\lambda$5007 emission than their radio-undetected counterparts.
These studies were, however, limited by the resolution of the FIRST and NVSS surveys, so that ``compact'' meant unresolved at about 5~kpc scales, and the sizes of the smallest sources could not be measured directly. \citet{escott26} used sub-arcsecond LOFAR images to overcome this limitation and show that AGN radio emission on kpc scales was associated with the incidence of [O~{\sc iii}] outflows. Other evidence for jet-driven shocks in radio galaxies includes the detection of strong H$_{2}$ rotational emission lines \citep{ogle10}. To date, however, there has only been one small sample of bright $z\sim 1$ 3C and 6C radio galaxies \citep{inskip02} and a few individual cases \citep[e.g. 3C326N,][]{leftley24} studied where there is evidence for shocks both from emission line widths and from emission line diagnostic ratios.

We were thus motivated to use the power of two new sky surveys to obtain a better understanding of the incidence and nature of feedback from radio jet powered shocks by
focusing on whether the size of a radio source is connected to the properties of the ionized gas in its host galaxy. The Very Large Array Sky Survey \citep[VLASS;][]{lacy2020} is the highest angular resolution radio survey to image the entire sky visible to the Very Large Array (VLA), achieving a physical resolution of $\approx 10$~kpc at $z=0.3$. The Dark Energy Spectroscopic Instrument (DESI) Data Release 1  \citep{DESI26}
provides high quality and wide spectral range optical spectroscopy of the host galaxies of many VLASS radio sources, allowing investigation of both the [O~{\sc iii}] emission line profiles and another potential indicator of AGN and shocks, namely the [O~{\sc i}]/H$\alpha$ vs [O~{\sc iii}]/H$\beta$ BPT diagram \citep{BPT, VO87, Allen08}.
From a VLASS and DESI cross-match we selected two samples of [O~{\sc i}]~$\lambda$6300 detected sources at $0.15 < z < 0.5$. The first contains extended sources, whose jets terminate outside the main body of the host. The second contains compact, but resolved sources whose jets are outside of the nuclear region, but still working their way through the ISM of the host.
We then examine the emission line properties of these sources, in particular the W80 distribution of the [O~{\sc iii}] line, their position in the BPT diagram and the presence or absence of the high ionization [Ne~{\sc v}]~$\lambda$3426 emission line in order to compare the role of shocks in ionizing the emission line gas between the two samples. We assume a cosmology with $H_0=70\,{\rm km\,s^{-1}\,Mpc^{-1}}$, $\Omega_{\rm M}=0.3$ and $\Omega_{\Lambda}=0.7$.

\section{Methods}

\subsection{Sample selection}

We used the 20260601 versions of the VLASS Single Epoch Continuum source catalogs from VLASS Epochs 2.1 and 3.2 available from the SE Continuum area of the VLASS data repository \footnote{\url{https://archive-new.nrao.edu/vlass}} to select the radio sources used in this study. These catalogs consist of both compact and extended (double or triple) sources, the latter being selected using the DRAGN Hunter algorithm of \citet{gordon23} (see \citet{kimball26} for more details of these catalogs).
The VLASS catalogs were then matched to the DESI data release 1 {\sc agngal} value-added catalog (Juneau et al., in preparation\footnote{\url{https://data.desi.lbl.gov/public/dr1/vac/dr1/agngal/v1.0/}}) using a 1.5 arcsec match radius. We then selected objects with [O~{\sc i}]~6300\,\AA\ detected at $>5\sigma$ and redshifts between 0.15 and 0.5 (the redshift range being determined so as to include both the high ionization [Ne~{\sc v}]~3426\,\AA\ line and the low ionization [O~{\sc i}]~6300\,\AA\ line, together with H$\alpha$, in the spectra).

For this study, we then selected two subsamples. One with compact radio sources that were nevertheless resolved in VLASS (so neither confined to the nuclear region nor dominated by a single strongly beamed blazar-like jet component), but likely to be contained within the host galaxy, and one where the radio jets were of sufficient extent that their working surfaces are likely to have passed beyond the ISM of the host galaxy. The beamsize of VLASS is $\approx 2.5$\,arcsec, corresponding to 6--15~kpc in the redshift range of the sample. We therefore selected sources with projected physical sizes between 10~kpc and 15~kpc in size for the ``small'' sample (although the lower limit of 10~kpc is below the beamsize at $z=0.5$, objects can be reliably resolved in the source catalogs to a limit of about half the beamsize, as the flux density cut for components is made at 1.2~mJy, $\approx 7$--$10\sigma$; \citet{kimball26}). The ``large'' subsample was selected to have a projected minimum size of 25~kpc. This limit is a compromise: it is large enough that, except for sources close to the line of sight, the jets terminate well outside the stellar body of the host, while being small enough that the two subsamples remain comparable in size. The gap between 15 and 25~kpc is deliberate, so that projection alone is unlikely to move a source from one subsample to the other. Each radio source and its putative identification were then examined by eye to ensure that the radio source was correctly classified as ``large'' or ``small'' and that its optical identification was correct. 
Finally, we noted that, as VLASS, like all interferometric surveys, is brightness limited, the fluxes and hence radio luminosities of our ``small'' and ``large'' samples differed, with the typical ``small'' object being $\approx 10\times$ fainter than the typical ``large'' one. To make the luminosity distributions of the sample more comparable and to eliminate the lowest luminosity sources we therefore placed a lower bound of 3~GHz luminosity of $>10^{24}$~W\,Hz$^{-1}$ on both samples, which also ensured they all fall into the traditional ``radio loud'' category \citep[e.g.][]{miller90}, with radio luminosities dominated by jet emission. As a check to ensure that the small sample was not unduly dominated by objects with jets close to the line of sight that might have much larger intrinsic sizes, we compared the fraction of type-1 AGN in each sample, taking a source to be type-1 when the {\sc agngal} catalog reports a nonzero broad H$\alpha$ flux. We find 13 of the 44 small sources and 18 of the 33 large sources are type-1, suggesting no such bias.

\subsection{Line measurements}

We downloaded the DESI spectra for all objects in both samples and fit the emission lines with sums of Gaussian components. The [O~{\sc iii}] $\lambda\lambda$4959,5007 doublet was fit with narrow and broad components for each line, with the doublet amplitude ratio fixed at the theoretical value. The [O~{\sc i}] $\lambda$6300 line was fit in the same way; the [O~{\sc i}] fit is used for the line ratio and luminosity only, with W80 measured from the [O~{\sc iii}] profile. For the BPT line ratios, H$\beta$ and H$\alpha$ were each fit with narrow, broad and very broad components (with the very broad component assumed to arise in the broad line region), so the two Balmer lines share the same kinematic structure. H$\alpha$ was fit simultaneously with the [N~{\sc ii}] $\lambda\lambda$6548,6583 doublet, which was given narrow and broad components with velocities and widths tied to those of H$\alpha$, and the doublet ratio fixed at 1/3 by atomic physics. A component was considered detected when its amplitude exceeded three times its fit uncertainty. The centroid of the broad component was initially restricted to within $\pm5$\,\AA\ (about $\pm230$~km~s$^{-1}$ at [O~{\sc iii}]) of the systemic velocity. For the broadest objects we repeated the fit with the centroid free, since in one case the constrained fit had settled at the edge of that range.

From the total fitted profile of each set of [O~{\sc iii}]~$\lambda\lambda$4959,5007 lines we measured W80, the velocity width containing the central 80 percent of the line flux \citep{liu2013}, corrected for the instrumental resolution. Because W80 is measured from the summed profile, it does not depend on how the flux is divided between the Gaussian components. Each object was classified as a Seyfert, a LINER, or star forming on the [O~{\sc i}]/H$\alpha$ versus [O~{\sc iii}]/H$\beta$ diagram using the boundaries of \citet{kewley2006}. After applying the luminosity cut of $L_{\rm 3\,GHz} > 10^{24}$~W~Hz$^{-1}$ the final samples contain 44 small and 33 large sources with all four lines measured.

In addition to the fitted lines, we made use of two further measurements. The first is the high ionization [Ne~{\sc v}]~$\lambda$3426 line. Its ionization potential of 97~eV is too high for stellar photoionization, and \citet{negus23} show that [Ne~{\sc v}] is an effective tracer of AGN activity, detected preferentially in more luminous AGN. We took the [Ne~{\sc v}] fluxes and their uncertainties from the {\sc agngal} catalog, and counted a source as detected when the flux exceeded three times its uncertainty. All 77 objects have spectral coverage of the line. The second is the Balmer decrement, the ratio of the narrow H$\alpha$ to H$\beta$ fluxes, which we use as an indicator of dust reddening: in the absence of dust the ratio is expected to be $\approx 3$ for AGN narrow-line regions \citep[e.g.][]{osterbrock89}, and dust raises it.

\section{Results}

\begin{figure*}
\plotone{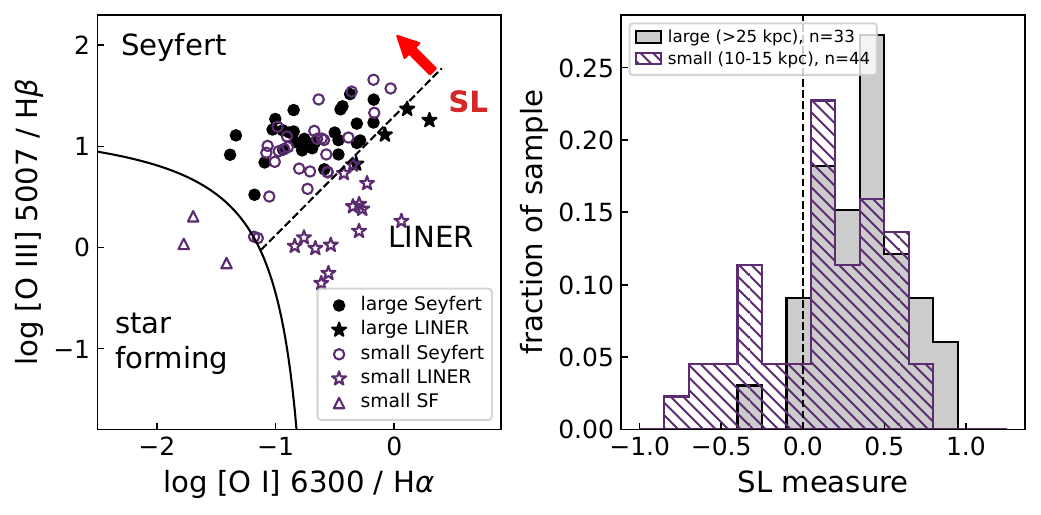}
\caption{\textbf{Left:} the two samples on the [O~{\sc i}]/H$\alpha$ versus [O~{\sc iii}]/H$\beta$ diagram, with the classification boundaries of \citet{kewley2006}. The large sample ($>25$~kpc) is almost entirely Seyfert, while the small sample (10--15~kpc) is a mix of Seyferts and LINERs. The arrow shows how the perpendicular distance from the Seyfert--LINER dividing line -- the SL (Seyfert--LINER) measure -- is defined. Right: histogram of the SL measure, with positive distances on the Seyfert side. The small sources sit systematically closer to the LINER boundary.}
\label{fig:bpt}
\end{figure*}

Figure~\ref{fig:bpt} shows the two samples on the [O~{\sc i}]/H$\alpha$ versus [O~{\sc iii}]/H$\beta$ BPT diagram. The two samples differ in their mix of AGN types. The large sample almost entirely falls within the region of the diagram occupied by Seyfert galaxies: 29 of 33 classified objects are Seyferts, four are LINERs, and none fall in the star-forming region. The small sample is mixed, with 27 Seyferts, 14 LINERs, and three objects that fall within the region occupied by star-forming galaxies (though inspection of the DESI spectra showed that all three of these are AGN, one is a type-2 AGN with strong [Ne~{\sc v}] emission and two are type-1 AGNs).

\begin{figure*}
\plotone{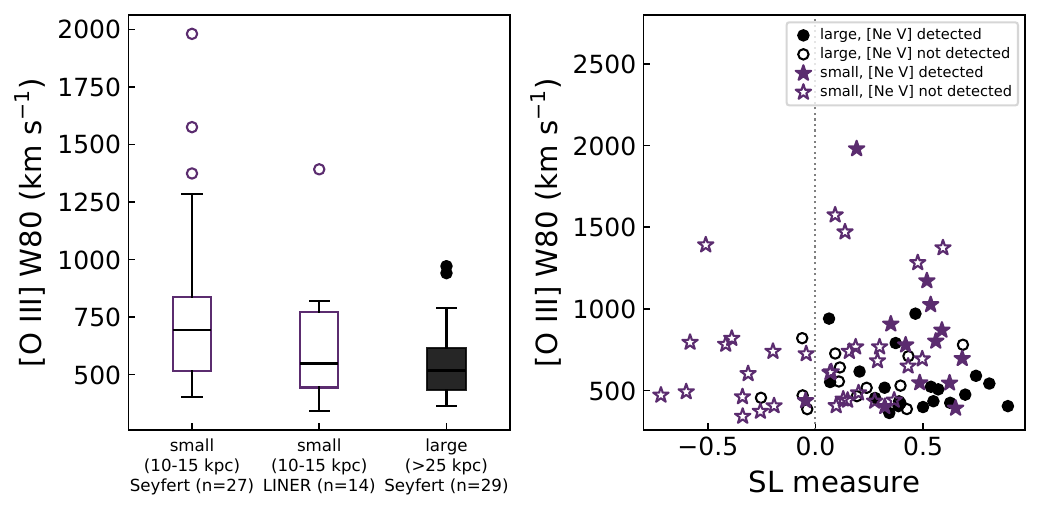}
\caption{\textbf{Left:} [O~{\sc iii}] W80 by radio size and AGN type. Small-sample Seyferts are significantly broader than large-sample Seyferts, while Seyferts and LINERs within the small sample are indistinguishable. The large-sample LINER bin contains too few objects to plot. {Right: W80 against the SL measure for all 77 sources, black circles for the large ($>25$~kpc) sample and purple stars for the small (10--15~kpc) sample; filled symbols are [Ne~{\sc v}]~$\lambda$3426 detections at $>3\sigma$ and open symbols non-detections. The dotted line marks the Seyfert--LINER divide. Of the five broadest small sources, only the broadest is detected in [Ne~{\sc v}]. No strong correlation is present within either sample.}}
\label{fig:w80}
\end{figure*}

Figure~\ref{fig:w80} shows the W80 distributions. The small sources have broader [O~{\sc iii}] profiles, and the extreme tail above 1300~km~s$^{-1}$ belongs entirely to the small sample. To separate the effect of the jet from the effect of the AGN type we compared the Seyferts alone. The {27} small-sample Seyferts have a median W80 of {694}~km~s$^{-1}$ against {518}~km~s$^{-1}$ for the {29} large-sample Seyferts, a difference that is significant on a Kolmogorov--Smirnov (KS) test ({$p = 0.020$}, where $p$ is the probability of the null hypothesis that the two distributions are from the same underlying sample) and a Mann--Whitney test ({$p = 0.0087$}). Within the small sample, the Seyferts and the LINERs show no detectable difference in W80 (medians {694} and 548~km~s$^{-1}$, KS {$p = 0.55$}), although the LINER bin is small. The line widths therefore track the size of the radio source rather than the AGN classification.

\begin{figure*}
\plotone{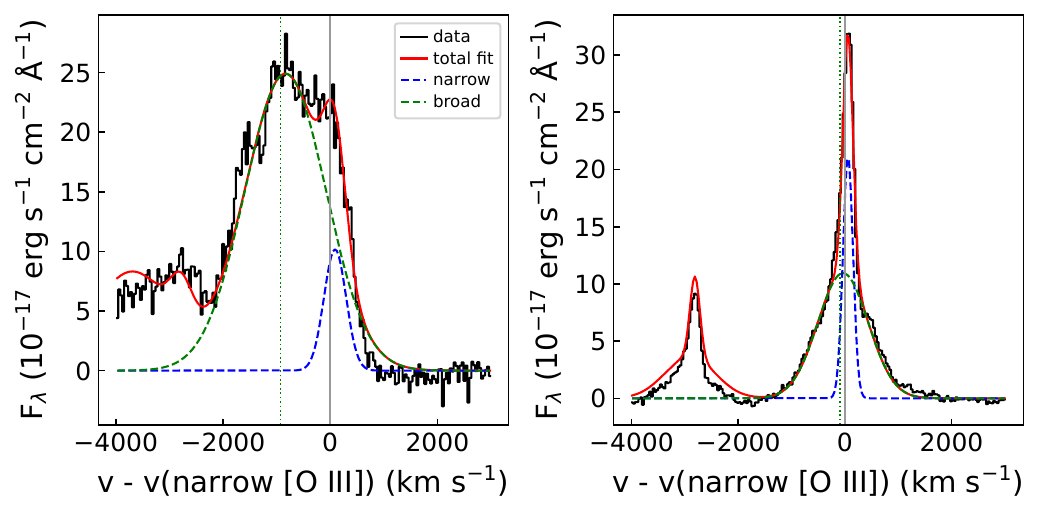}
\caption{{[O~{\sc iii}]~$\lambda\lambda$4959,5007 profiles of two small-sample sources, in velocity relative to the narrow [O~{\sc iii}] component. Data in black; total fit in red; narrow and broad components dashed. Left: VLASS3.2SE J161431.68+152421.3, the broadest source in the sample, with the broad component blueshifted by $925\pm13$~km~s$^{-1}$ and W80 $=1981$~km~s$^{-1}$. Right: VLASS2.1SE J172206.08+565452.1, a source with W80 $=1026$~km~s$^{-1}$ whose broad component is offset by only $-93$~km~s$^{-1}$. The secondary peak near $-2900$~km~s$^{-1}$ in the right panel is [O~{\sc iii}]~$\lambda$4959; in the left panel the lines are broad enough that the doublet blends.}}
\label{fig:profile}
\end{figure*}

{A broad [O~{\sc iii}] profile does not by itself distinguish shocked gas from an AGN-driven outflow viewed at a favourable angle. The velocity of the broad component offers one handle: an outflow whose receding side is hidden by the host should appear blueshifted, whereas gas stirred on both sides of a jet need not be. Figure~\ref{fig:profile} shows the profile of the broadest source in the small sample, with W80 $=1981$~km~s$^{-1}$. It has a narrow core within 100~km~s$^{-1}$ of the systemic velocity and a broad component blueshifted by $925 \pm 13$~km~s$^{-1}$, and is the most outflow-like profile in either sample. Among the six small sources with W80 above 1000~km~s$^{-1}$ and a detected broad component, however, only two have that component blueshifted by more than 150~km~s$^{-1}$; the other four lie within 150~km~s$^{-1}$ of systemic. The median broad-component offset is $-41$~km~s$^{-1}$ in the small sample and $-57$~km~s$^{-1}$ in the large one, so the two samples differ in the width of their lines rather than in a systematic velocity shift. An outflow contribution in individual objects cannot be excluded, and the object in the left panel of Figure~\ref{fig:profile} is a natural candidate.}

To test the significance of the separation on the BPT diagram, we computed, for every object, the perpendicular distance to the Seyfert--LINER dividing line of \citet{kewley2006}, with positive distances on the Seyfert side (Figure~\ref{fig:bpt}, right), which we refer to as the ``SL measure''. The large sample has a median SL measure of {0.39} against {0.17} for the small sample. The difference is significant on a Mann--Whitney test ($p = 0.020$) and an Anderson--Darling test ($p = 0.017$), but not on a KS test ($p = 0.10$), which is less sensitive to a difference concentrated in one tail of the distributions.

We next used the [Ne~{\sc v}] detection rate as a diagnostic of the ionizing spectrum rather than as an AGN identifier. As noted above, [Ne~{\sc v}] traces gas photoionized by the UV continuum of an AGN accretion disk. In gas ionized by shocks, however, the shock velocities need to be high, $\gtrsim 600$~km~s$^{-1}$, before large amounts of [Ne~{\sc v}] are produced \citep{Allen08}. We therefore compared the [Ne~{\sc v}] detection rates of the large and small samples. In the large sample {58} per cent of sources are detected at $>3\sigma$, against {30} per cent of the small sample {(Fisher exact $p = 0.019$)}; at $>5\sigma$ the rates are {55} and {20} per cent {($p = 0.003$)}. Restricting both samples to a matched redshift range of $0.30 < z < 0.45$ ({19 small and 21 large}) {gives detection rates similar to those of the full samples ({57 against 37} per cent), although with about 20 objects in each the difference is not significant on its own ($p = 0.22$), so redshift does not appear to drive it. The median [Ne~{\sc v}] flux uncertainties are nearly identical, but because the small sources are fainter in [O~{\sc iii}], their upper limits on [Ne~{\sc v}]/[O~{\sc iii}] are about twice as shallow. Among the detections alone, the small sources still have the lower [Ne~{\sc v}]/[O~{\sc iii}] ratio (medians {0.051 and 0.089}, Mann--Whitney $p = 0.019$).}

One further possibility is dust, since [Ne~{\sc v}] lies in the blue where extinction is strongest. The small sample is slightly dustier, with a median H$\alpha$/H$\beta$ of {4.01} against {3.64} for the large sample (Mann--Whitney {$p = 0.026$}). This modest difference cannot account for the [Ne~{\sc v}] gap, however: restricting both samples to H$\alpha$/H$\beta < 4$ widens the gap to {76 against 36} per cent {($p = 0.014$)}. Within the large sample the [Ne~{\sc v}] non-detections do have higher decrements than the detections (medians {4.13 against 3.26}, $p = 0.01$), but no such difference exists in the small sample ({$p = 0.22$}). The lower [Ne~{\sc v}] detection rate in the small sources therefore appears to reflect a real difference in the ionizing conditions rather than dust.

{The strength of [Ne~{\sc v}] depends on the hardness of the ionizing continuum as well as on its presence, so we asked whether the two samples might simply host AGN with different continua. Among the Seyferts only, which removes most of the difference in optical class, the detection rates are {37} per cent (small) and {66} per cent (large; {$p = 0.06$}). Restricting to Seyferts with [O~{\sc iii}]/H$\beta > 5$, a rough cut on the hardness of the ionizing spectrum, narrows the gap to 56 against 68 per cent ($p = 0.52$), and matching the samples one to one in [O~{\sc iii}] luminosity (within 0.15~dex, 26 pairs) narrows it to 35 against 50 per cent, so some of the raw difference does follow luminosity and excitation. A smaller gap among Seyferts alone is expected if part of the difference reflects the larger shock contribution in the LINER-like sources.}

{If shocks are responsible for the broad lines and only shocks faster than $\approx600$~km~s$^{-1}$ produce much [Ne~{\sc v}], the objects with the largest W80 within a sample might be expected to be the ones detected in [Ne~{\sc v}]. Within the small sample the W80 distributions of the [Ne~{\sc v}] detections and non-detections are indistinguishable (Mann--Whitney $p = 0.41$), as they are within the large sample ($p = 0.57$). The numbers are small, and W80 from an integrated fibre spectrum is at best a loose guide to the velocity of the shock behind any one line, but the prediction of the simplest velocity-ordered picture is not borne out within either sample.}

\section{Discussion}
All three measurements point the same way. The small radio sources, whose jets are still inside the host, have on average broader [O~{\sc iii}] profiles, sit closer to the LINER region of the BPT diagram and tend to lack [Ne~{\sc v}] emission, while the large sources, whose jets terminate outside the host, have narrower lines, frequently have detectable [Ne~{\sc v}] and are almost uniformly Seyferts. LINER-like line ratios can originate from either shocks \citep[both through direct shock heating and by photoionization by post-shock gas e.g.][]{Allen08}, or from hard ionizing spectra directly from an AGN \citep[e.g.][]{kewley2006}. In this case, however, the association with radio jets within the host galaxy and the lack of [Ne~{\sc v}] emission point strongly towards a shock origin. {The shape of the accretion disk continuum is set on scales far smaller than the radio source, and there is no obvious reason for it to depend on the projected size of the jets except through the age of the source. A systematic difference in continuum hardness between the samples would therefore need its own explanation, whereas a larger shock contribution to the line emission in the confined sources follows directly from the size selection. Selecting on [O~{\sc i}] does favour gas-rich hosts with extended partially ionized regions, but that requirement is applied identically to both samples and is independent of radio size, and the samples differ not only in [O~{\sc i}]/H$\alpha$ but also in [O~{\sc iii}]/H$\beta$, in line width and in [Ne~{\sc v}] detectability, none of which the [O~{\sc i}] cut acts on, so the selection cannot be responsible for the differences we find.} These results therefore favor the jets of the small sources driving shocks through the interstellar medium of their hosts and powering much of the line emission. There are, however, some small sources that are firmly in the Seyfert region of the BPT diagram, with a larger distance ($\gtrsim +0.3$) from the Seyfert-LINER divide than the shock models of \citet{Allen08} predict. In these cases we may be seeing AGN photoionization dominating even for objects with high W80.

{The line ratios used for classification come from the total fitted profiles, so we also asked whether the individual kinematic components fall in different parts of the diagram. [O~{\sc i}] is too faint to decompose in most objects, so the components cannot be placed on the [O~{\sc i}]/H$\alpha$ axis. In [O~{\sc iii}]/H$\beta$, measurable per component in 15 objects, the broad component has the lower ratio in 11, that is, it lies nearer the LINER region than the narrow component. This holds in both samples and so reflects the broad gas in general rather than jet confinement. Total fluxes do blend gas of different excitation, but the narrow components alone still separate the samples in the same sense (median $\log$[O~{\sc iii}]/H$\beta$ of 0.75 against 0.98), though not significantly given the smaller numbers ($p = 0.13$), so we do not believe the classifications in Figure~\ref{fig:bpt} are a product of that blending.}

In contrast to the majority of the small sources, the large sources lack shock signatures, consistent with the gas around the large sources having relaxed after the jet broke out of the galaxy, with the emission line gas now being photoionized principally by UV continuum emission from the accretion disk of the AGN. After breakout the jet working surfaces drive shocks into the much lower density circumgalactic medium, where cooling times are long and the shocks are no longer fully radiative \citep[e.g.][]{dopita03}, so that little optical line emission is expected from them even while the jet remains active.

To ask how much of the jet kinetic power ends up radiated in the emission lines, we estimated the bulk kinetic power $Q$ of the jets following \citet{willott1999}, who calibrated $Q$ against the 151~MHz luminosity of FR~II radio sources using minimum-energy arguments together with a model for how a radio source grows into its surrounding medium,
\begin{equation}
Q \simeq 3\times10^{38} \, f^{3/2}
\left(\frac{L_{151}}{10^{28}~{\rm W\,Hz^{-1}\,sr^{-1}}}\right)^{6/7}~{\rm W},
\end{equation}
where $f$ collects the uncertainties in the jet composition, the volume filling factor of the lobe plasma, the low-energy cut-off of the electron energy distribution, and any departure from the minimum-energy condition. \citet{willott1999} argue that $f$ lies between 1 and 20. We adopt $f=1$, so the values of $Q$ quoted here are lower limits; adopting $f=20$ would raise them by a factor of $\approx 90$, consistent with the study of \citet{cavagnolo10}. The relation is calibrated on FR~II sources and assumes a simple model for the surrounding density, so may not be appropriate for our confined sources, but we believe it is useful as an order of magnitude guide to the jet power. We converted each VLASS 3~GHz luminosity to 151~MHz assuming a spectral index of $\alpha = -0.7$.
The median jet power is {$\log_{10}(Q/{\rm W}) = 35.7$} for the large sample and 35.1 for the small sample.

\begin{figure}
\centering
\includegraphics[width=3.5in]{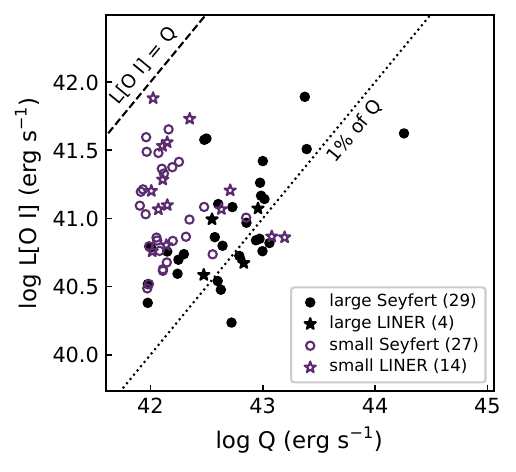}
\caption{[O~{\sc i}] luminosity against jet kinetic power $Q$, estimated from the radio luminosity following \citet{willott1999} with $f=1$. Filled black points are the large ($>25$~kpc) sample and open purple points the small (10--15~kpc) sample; circles are Seyferts and stars are LINERs; the three small-sample objects in the star-forming region are not plotted. The dashed and dotted lines mark $L_{\rm [O\,I]} = Q$ and $L_{\rm [O\,I]} = 0.01\,Q$. $Q$ is plotted in erg~s$^{-1}$ (1~W $= 10^{7}$~erg~s$^{-1}$).}
\label{fig:loiq}
\end{figure}

Figure~\ref{fig:loiq} compares the [O~{\sc i}] luminosity with the jet power. Most sources radiate more than one per cent of their jet power in [O~{\sc i}] alone, and the least powerful sources, which are confined, {reach several tens of per cent, up to 0.7 in one case}. Since [O~{\sc i}] is only a minor coolant, the total emission-line luminosity would then be an implausibly large fraction of $Q$, which suggests that $f=1$ is indeed a conservative lower limit on the jet power. Within each sample the Seyferts and LINERs occupy the same region of this diagram, consistent with the W80 results in that the radio source size, rather than the AGN classification, is what distinguishes the two samples.

The right-hand panel of Figure~\ref{fig:w80} compares W80 directly against the SL measure. No strong correlation is present within either sample, which is not unexpected if a mix of shock heating and both AGN and shock photoionization contributes to the line emission in individual objects.

\begin{figure*}
\plotone{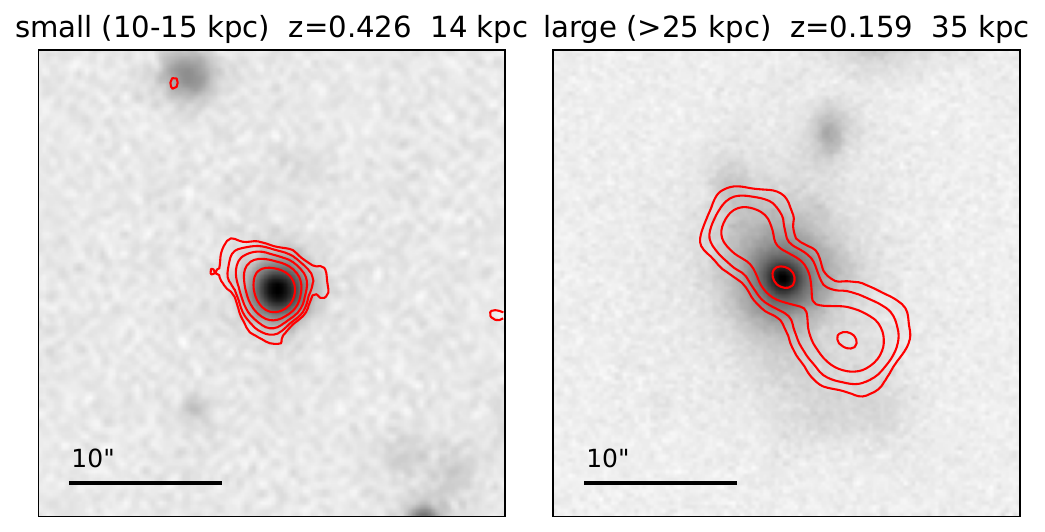}
\caption{Two example sources from our study, shown on Legacy Surveys \citep{dey19} DR10 $r$-band images $31\farcs4$ on a side, with VLASS Epoch 1.2 Quick Look contours at 3, 6, 12, 24 and 48 times the rms noise of each image ($\sigma = 0.16$ and $0.23$~mJy~beam$^{-1}$ for the left and right panels). Left: VLASS2.1SE J172206.08+565452.1, from the small (10--15~kpc) sample. Right: VDGN2.1SE J141613.25+021906.4, from the large ($>25$~kpc) sample. The bar marks $10''$.}
\label{fig:examples}
\end{figure*}

We also checked whether there was any difference in the host galaxy populations between the large and small samples by inspecting images from the Legacy Survey \citep{dey19} DR10 of all 77 hosts. Every host is consistent with an elliptical or a merging system. The merger fraction does not differ significantly between the samples: 26 of 44 (59 per cent) small and 15 of 33 (45 per cent) large hosts show signs of interaction. Figure~\ref{fig:examples} shows one host from each sample with the VLASS contours overlaid.

\section{Conclusions}
We compared the ionized gas in two samples of radio-loud AGN selected from VLASS and DESI DR1: {44} sources with 10--15~kpc jets still confined within their hosts, and {33} sources with $>25$~kpc jets that have broken out. The small sample has a somewhat lower median jet power than the large sample, so the differences we find cannot be attributed to more powerful jets in the confined sources.

We found three differences between the samples. Among the Seyferts alone, the confined sources have broader [O~{\sc iii}] profiles, with no corresponding systematic blueshift of the broad component. The confined sources also sit closer to or below the LINER boundary. Finally, we detect [Ne~{\sc v}] in the confined sources only half as often. This difference is not explained by redshift or dust and is present among the Seyferts alone, if at reduced significance; restricting in [O~{\sc iii}] luminosity or in [O~{\sc iii}]/H$\beta$ narrows it, so part of it reflects a difference in the mean AGN properties between the samples.

These differences indicate different ionizing mechanisms at work in the two samples. In the sources whose jets have broken out the accretion disk is mostly responsible, since the line ratios are almost uniformly Seyfert-like and the frequent [Ne~{\sc v}] detections show that the hard UV continuum of the accretion disk reaches the line-emitting gas. In the confined sources the evidence points to significant contributions from shocks driven by the jet. No single measurement establishes this on its own, since broad lines do not identify their driver and LINER-like ratios can also be produced by a hard but dilute AGN radiation field. It is the combination that is decisive, because the LINER-like ratios and the disturbed kinematics appear specifically in the sources whose jets are still inside the host, and the [Ne~{\sc v}] deficit shows that the disk radiation is not the dominant ionizing source for much of their gas, which disfavors the dilute AGN field alternative.
The lack of correlation between W80 and BPT distance within the confined sample is consistent with the relative contributions of shock heating, shock photoionization and AGN photoionization varying from object to object.
The [O~{\sc i}] luminosities, which exceed one per cent of the jet kinetic power in most sources, show that the jets carry more than enough energy to power the line emission.
Taken together, our results indicate that jets confined within the host drive shocks through the interstellar medium and power the bulk of the line emission of most of the small sample, while the gas in the escaped sources has relaxed and is photoionized principally by the AGN accretion disk.

The data underlying this study, ``Data for: Evidence for strong shocks in the ISM of small radio sources'', are available on Zenodo under a CC BY 4.0 license at \url{https://doi.org/10.5281/zenodo.23326653}. The scripts that produce the figures are at \url{https://github.com/valelop900/lopez26_shocks}.

\begin{acknowledgments}
We thank Kristina Nyland and Pallavi Patil for helpful discussions and the referee for constructive comments. VL thanks the NRAO student program for support. The National Radio Astronomy Observatory is a facility of the National Science Foundation operated under cooperative agreement by Associated Universities, Inc. This research used data obtained with the Dark Energy Spectroscopic Instrument (DESI). DESI construction and operations is managed by the Lawrence Berkeley National Laboratory. This material is based upon work supported by the U.S. Department of Energy, Office of Science, Office of High-Energy Physics, under Contract No. DE–AC02–05CH11231, and by the National Energy Research Scientific Computing Center, a DOE Office of Science User Facility under the same contract. Additional support for DESI was provided by the U.S. National Science Foundation (NSF), Division of Astronomical Sciences under Contract No. AST-0950945 to the NSF’s National Optical-Infrared Astronomy Research Laboratory; the Science and Technology Facilities Council of the United Kingdom; the Gordon and Betty Moore Foundation; the Heising-Simons Foundation; the French Alternative Energies and Atomic Energy Commission (CEA); the National Council of Humanities, Science and Technology of Mexico (CONAHCYT); the Ministry of Science and Innovation of Spain (MICINN), and by the DESI Member Institutions: www.desi.lbl.gov/collaborating-institutions. The DESI collaboration is honored to be permitted to conduct scientific research on I’oligam Du’ag (Kitt Peak), a mountain with particular significance to the Tohono O’odham Nation. Any opinions, findings, and conclusions or recommendations expressed in this material are those of the author(s) and do not necessarily reflect the views of the U.S. National Science Foundation, the U.S. Department of Energy, or any of the listed funding agencies.
\end{acknowledgments}

\begin{contribution}

Both authors contributed equally to this paper.


\end{contribution}

%
\facilities{VLA, Mayall(DESI)}

\software{astropy \citep{astropy13,astropy18,astropy22}, matplotlib \citep{hunter07},
          TOPCAT \citep{taylor05}, pyspeckit \citep{ginsburg22},
          scipy \citep{virtanen20}, numpy \citep{harris20}}

\bibliography{sample701}{}
\bibliographystyle{aasjournalv7}

\end{document}